# Room-temperature ferroelectric switching of spin-to-charge conversion in GeTe

Sara Varotto[1,*], Luca Nessi[1,2], Stefano Cecchi[3], Jagoda Sławińska[4,5], Paul Noël[6], Simone Petrò[1], Federico Fagiani[1], Alessandro Novati[1], Matteo Cantoni[1,2], Daniela Petti[1], Edoardo Albisetti[1], Marcio Costa[7], Raffaella Calarco[3,8], Marco Buongiorno Nardelli[4], Manuel Bibes[9], Silvia Picozzi[10], Jean-Philippe Attané[6], Laurent Vila[6], Riccardo Bertacco[1,2], Christian Rinaldi[1,2†]

[1]*Dipartimento di Fisica, Politecnico di Milano, via G. Colombo 81, 20133 Milan, Italy*
[2]*Istituto di Fotonica e Nanotecnologie IFN-CNR, c/o piazza Leonardo da Vinci 32, 20133 Milan, Italy*
[3]*Paul-Drude-Institut für Festkörperelektronik, Hausvogteiplatz 5-7, 10117 Berlin, Germany*
[4]*Department of Physics, University of North Texas, Denton, TX 76203, United States of America*
[5]*Zernike Institute for Advanced Materials, University of Groningen, Nijenborgh 4, 9747AG Groningen, Netherlands*
[6]*Université Grenoble Alpes, CEA, CNRS, Grenoble INP, Spintec, 38000 Grenoble, France*
[7]*Department of Physics, Fluminense Federal University, 24210-346, Niterói, Rio de Janeiro, Brazil*
[8]*Consiglio Nazionale delle Ricerche - CNR-IMM, Via del Fosso del Cavaliere 100, 00133 Rome, Italy*
[9]*Unité Mixte de Physique, CNRS, Thales, Université Paris-Saclay, 91767 Palaiseau, France*
[10]*Consiglio Nazionale delle Ricerche, CNR-SPIN c/o Università G. D'Annunzio, 66100 Chieti, Italy*

---

[*] Present address: *Unité Mixte de Physique, CNRS, Thales, Université Paris-Saclay, 91767 Palaiseau, France*
[†] E-mail: christian.rinaldi@polimi.it

# Abstract

Since its birth in the 1990s, semiconductor spintronics has suffered from poor compatibility with ferromagnets as sources of spin. While the broken inversion symmetry of some semiconductors may alternatively allow for spin-charge interconversion, its control by electric fields is volatile. Ferroelectric Rashba semiconductors stand as appealing materials unifying semiconductivity, large spin-orbit coupling, and non-volatility endowed by ferroelectricity. However, their potential for spintronics has been little explored. Here, we demonstrate the non-volatile, ferroelectric control of spin-to-charge conversion at room temperature in epitaxial GeTe films. We show that ferroelectric switching by electrical gating is possible in GeTe despite its high carrier density. We reveal a spin-to-charge conversion as effective as in Pt, but whose sign is controlled by the orientation of the ferroelectric polarization. The comparison between theoretical and experimental data suggests that spin Hall effect plays a major role for switchable conversion. These results open a route towards devices combining spin-based logic and memory integrated into a silicon-compatible material.

While current information and communication technology is based on semiconductors, new approaches such as spin-based electronics – or spintronics – are emerging as possible low-power solutions for beyond-CMOS. In an ideal scenario that could limit the need to develop and integrate exotic materials into the established CMOS platform, spin would be added as an extra degree of freedom into semiconductor-based electronics that operate on charge. The concept of semiconductor spintronics[1,2] emerged in the 1990s, propelled by the very long quantum coherence of spins in semiconductors and the appeal of integrating spintronics with optoelectronics. However, combining ferromagnets as spin generators and detectors with semiconductors proved a daunting endeavour thwarted by materials compatibility and impedance mismatch issues[3]. An alternative approach relying on doping magnetic impurities into semiconductors (as for instance in (Ga,Mn)As) was quite successful but could not reproducibly yield ferromagnetism at room temperature[4], despite a recent report on iron-doped semiconductors[5].

Another remarkable feature of zinc-blende semiconductors (such as GaAs) and their heterostructures is the broken inversion symmetry. It leads to the splitting of some of the electronic bands by the spin-orbit interaction, coupling the direction of spins and momentum and allowing for the Dresselhaus and Rashba effects. This spin-momentum locking, as well as the spin Hall effect, enable the generation of spin currents by charge currents and vice versa, so that semiconductor heterostructures can be used as spin generators and detectors[6–9]. However, at variance with ferromagnets, they lack the reversible non-volatile control of the spin, whose direction is either fixed by the construction of the heterostructure or set by internal electric fields, and thus volatile. Finally, their spin-charge interconversion process is not very efficient. Recently, some of us showed that the spin-to-charge conversion could be tuned by an electric-field-induced ferroelectric-like state in $SrTiO_3$ interfaces[10]. However, that system only works at low temperatures (< 50 K) and is difficult to integrate on silicon.

In this framework, the recently identified class of ferroelectric Rashba semiconductors (FERSC)[11] emerges as an exciting alternative to develop semiconductor devices operating on spin and integrating logic and memory functionalities. FERSC have a broken inversion symmetry just as several other semiconductors, but they are also ferroelectric. As such, they display a giant Rashba spin-splitting of the bulk bands, with the bonus that the spin direction in each Rashba sub-band can be reversed by switching the ferroelectric polarization[12]; our spin- and angle-resolved photoemission spectroscopy measurements have confirmed that the chirality of the spin texture is linked to the electric dipole in the FERSC prototype GeTe[13,14]. FERSC thus have a strong potential for the development of all-in-one devices integrating spin generation, manipulation and detection, similar to spin Hall effect transistors or magneto-electric spin orbit devices[15]. They are endowed with reconfigurability and non-volatility, just as ferromagnets but operating with electric fields, thereby offering a much-reduced power consumption. Moreover, the compatibility of chalcogenides with Ge[16] and GaAs[17] may open up the possibility to bridge photonics and spintronics (e.g. spin lasers) by exploiting the ferroelectric control of spin injection into optically active semiconductors.

Here we provide a direct evidence of a switchable spin-to-charge conversion (SCC) in epitaxial Fe/GeTe heterostructures at room temperature. First, we demonstrate the fast ferroelectric switching of GeTe thin films through gate electrodes with voltages compatibles with standard electronics. Moreover, we identify an efficient, non-destructive readout of the ferroelectric state. Then, by spin pumping experiments we reveal that the sign of the spin-to-charge conversion reverses when switching the electric polarization. Finally, we employ first principles calculations to shed more light on the origin of the observed phenomenon and reveal a major role of the inverse spin Hall effect in the SCC mechanism. Importantly, the spin-to-charge conversion in Fe/GeTe involves the bulk bands rather than the usually studied interface or surface Rashba states[18], thus distinguishing this system from 2D alternatives and opening the way to beyond-CMOS applications, with the monolithic integrability on silicon exploitable for the development of transistors as well as spin interconnects through the substrate[19].

## Ferroelectric switching of GeTe

The ferroelectric switching of GeTe has been demonstrated only at the nanoscale, by piezoresponse force microscopy (PFM) for thin films[14,20], transmission electron microscopy (TEM) for nanometre-scale crystals[21] and nanowires[22], and finally using X-ray spectroscopy[23]. The demonstration of the macroscopic gating of GeTe thin films has never been achieved. The common understanding of ferroelectric switching suggests that the screening of the external electric field by the large density of free carriers, would prevent the penetration of external electric fields, inhibiting polarization reversal and resulting in high dielectric loss[20]. Although very recent works have shown evidence of ferroelectric inversion even in polar metals[24–26], it was achieved only in the ultrathin limit.

From a technical perspective, the measurement of P-E loops via the positive-up negative-down method[27] commonly employed for ferroelectric oxides[28] cannot be performed. This is because GeTe is a semiconductor with a small indirect gap ($E_g$= 0.61-0.66 eV[29]), with high intrinsic p-type doping ($p\sim 10^{20}$ cm$^{-3}$) and relatively high conductivity ($\sigma\sim 5\times10^3$ cm$^{-1}$ $\Omega^{-1}$), so that displacement current peaks associated to the polarization switching are overshadowed by the conduction current. Therefore, a different method for the electric readout of the ferroelectric state must be identified for narrow-gap highly-doped semiconductors.

Aiming to demonstrate the gating, we performed a combined electric and piezoelectric experiment, following an approach similar to that used for FE oxides in Ref. [30]. Voltage pulses were applied by a source measure unit (SMU) to a metallic gate, deposited on the semiconductor and thin enough to permit the imaging of out-of-plane ferroelectric domains by PFM. A sketch of the device is shown in Fig. 1a.

After the application of a pulse $V_{\text{write}}$, we recorded the resistance of the heterojunction at low voltage, and then disconnected the SMU from the gate. A portion of the gate was scanned by the PFM tip excited with an *ac* signal of ~2 V at a frequency of ~70 kHz. The tip deflection demodulated by a lock-in amplifier, proportional to the piezo/ferroelectric response of the area under the tip, was collected to map the ferroelectric domains below the gate. Figure 1b reports such domains configuration that progressively evolves with $V_{write}$. Moving from +8 V to –8 V, it turns from a prevalent inward ($\mathbf{P}_{in}$) to a fully outward ($\mathbf{P}_{out}$) configuration, passing through intermediate states. The reversal was not complete for the positive pulse at +8 V (panel $b_1$), suggesting that outward domains were favoured and that a uniform inward state would have possibly required the application of higher voltages or longer pulses.

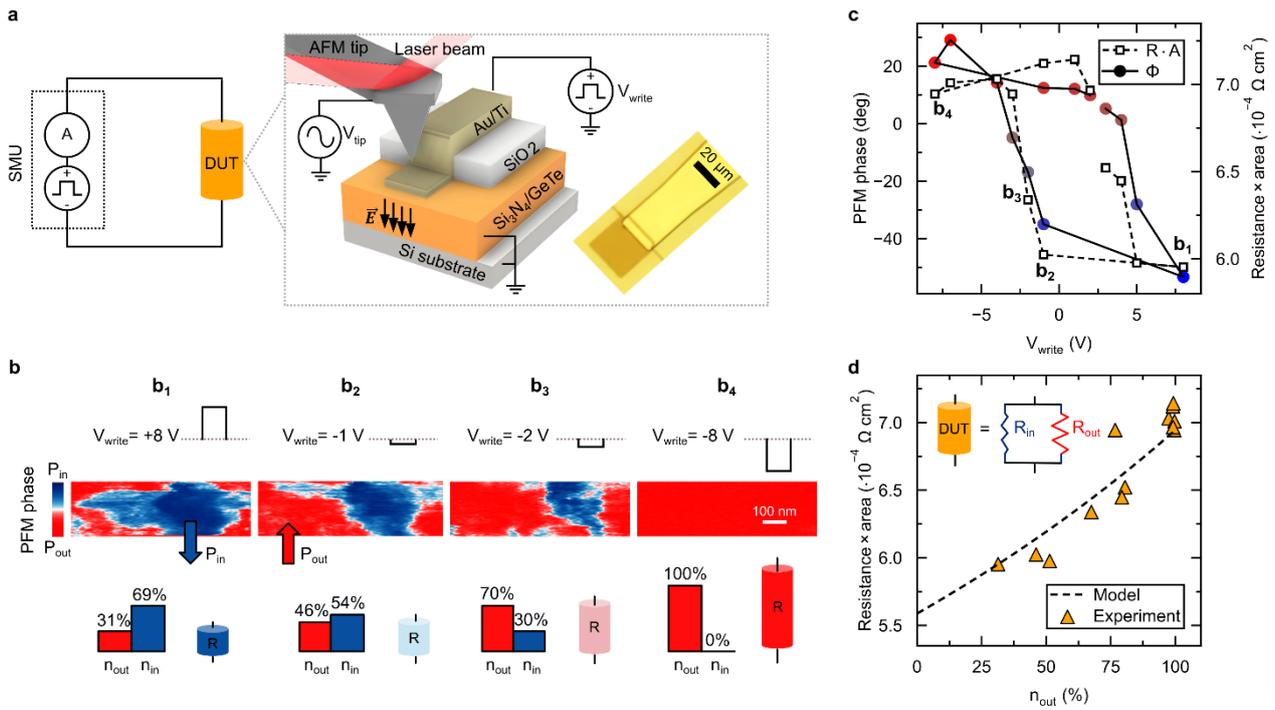

**Figure 1 | Ferroelectric switching of GeTe by electrical gating and electric readout of the state. a,** Sketch of the setup and the device under test (DUT) used for gating and ferroelectric imaging. Thin Au(5 nm)/Ti (5 nm) square gates (30x30 µm) were evaporated on $Si_3N_4$(10 nm)/GeTe(50 nm)/Si and connected to thicker Au(200 nm)/Cr(7 nm) stripes allowing for electrical contact with a SMU. An optical image of the device is reported on the right. **b₁- b₄,** PFM maps measured at remanence on the surface of the gate (scan size 0.7x0.2 µm) for $V_{write}$ from +8 V down to -8 V. The blue (red) state represents $\mathbf{P}_{in}$ ($\mathbf{P}_{out}$). The histograms with the percentage of outward ($n_{out}$) and inward ($n_{in}$) domains are reported below each map. **c,** PFM phase (circular dots) and RA product of the junction (black empty squares) versus the amplitude of the writing voltage pulse. The similarity of the two curves is a signature of ferroelectricity-driven resistive switching. Points $b_1$ ($V_{pulse}$ = +8 V), $b_2$ ($V_{pulse}$ = -1 V), $b_3$ ($V_{pulse}$ = -2 V), $b_4$ ($V_{pulse}$ = -8 V) correspond to the mean value of the PFM phase over the images reported in relative panels. **d,** Resistance-area product versus percentage of outward domains extracted from the PFM maps of panel b. Triangles refer to experimental data, while the dashed line is the trend expected from the parallel resistance model.

The mean value of the PFM phase calculated over the entire *xy* map for each value of $V_{write}$ is reported in Fig. 1c. It shows a hysteretic behaviour comparable to ferroelectric loops acquired on the bare surface, with bistability and a coercivity of about 3.5 V. An asymmetry of the switching is visible as a slight shift of the hysteresis loop (1.2 V), compatible with the different work functions of the electrodes or with built-in fields at GeTe interfaces that favour the preferential orientation of outward domains[31]. Figure 1c also reports the resistance–area (RA) product of the gate/Si$_3$N$_4$/GeTe junction, where the resistance $R$ is measured after each writing pulse at low voltage (0.1 V), while $A$ is the area of the electrode. The marked correlation with the PFM phase suggests that $R$ is connected to the ferroelectric state. To better understand this relationship, in Fig. 1d we report $R$ versus the relative fraction of outward domains $n_{out}$. The results can be interpreted using a simple model which assumes that $\mathbf{P}_{out}$ and $\mathbf{P}_{in}$ regions have different specific resistances ($R_{out}$ and $R_{in}$) and conduct the current in parallel[30,32], so that $R$ is

$$\frac{1}{R} = \left( \frac{n_{out}}{R_{out}} + \frac{1-n_{out}}{R_{in}} \right), \tag{1}$$

Such a model reproduces the experimental data (dashed line in Fig. 1d), corroborating the ferroelectric origin of the resistance modulation and providing an efficient method for the readout of the ferroelectric state.

We ascribe resistance changes to the modulation of the width of the Schottky barrier expected at the gate/semiconductor (SC) interface (inset of Fig. 2a). In the model developed by P. W. M. Blom *et al.*[33] for ferroelectric Schottky diodes, the height of the Schottky barrier $\Phi_{SB}$ is fixed and determined by the work function of the metal ($\phi_m$), the electron affinity of the semiconductor ($\chi_{sc}$) and its bandgap $E_g$ ($\Phi_{SB}= E_g+\chi_{sc}-\phi_m$)[34]. In this framework, the width of the depletion region depends on the effective dielectric constant of the semiconductor, whose value is determined by the volume polarization charges associated to a certain ferroelectric state. Self-consistent calculations[33] demonstrated that a narrower (or wider) depletion region is produced for a polarization $\mathbf{P}$ parallel (or anti-parallel) to the built-in electric field[33] inside the space-charge region, determining a lower (or higher) electrical resistance of the junction.

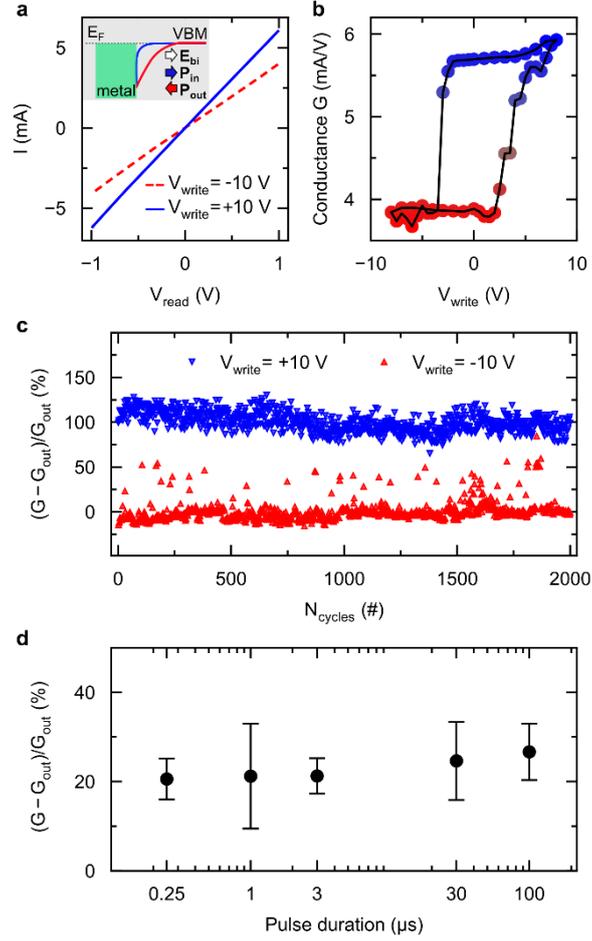

**Figure 2 | Resistive modulation in Ti/GeTe junctions. a,** Modulation of the current-voltage characteristic of the Ti/GeTe contact upon poling. A rectangular voltage pulse of 10 ms and amplitude $V_{write}$ sets the resistive state. The *I-V* curves are measured at relatively low voltages ($|V_{max}|= 1$ V) after two pulses at $V_{write}$= -10 V (red dashed line) and +10 V (blue solid line). The inset shows the corresponding qualitative modification of the band bending expected at the metal/GeTe interface, that depends on the relative orientation of polarization and built-in field $E_{bi}$. **b,** Conductance versus $V_{write}$ (calculated at $V_{max}$), showing a hysteresis loop compatible with those of ferroelectricity acquired by PFM on the free GeTe surface. **c**, Cyclability of the ferroelectric state measured by alternating negative and positive saturating pulses and measuring the conductance at low voltage at each step. **d,** Modulation of the junction conductance versus pulse duration. The standard error is used for the error bars.

We tested this mechanism for metal/GeTe junctions: arrays of squared Ti(100 nm) contacts were deposited on 22 nm thick GeTe(111) on Si(111). Voltage pulses were applied at the electrode to pole the ferroelectric GeTe film. Figure 2a reports the *I-V* curves collected at low voltages after poling with ±10 V. Due to the high *p*-type intrinsic doping of GeTe[35], the metal/semiconductor junction behaves as an ohmic contact (the Schottky barrier width is of the order of a few nanometres). The conductance is hysteretic (Fig. 2b), showing a modulation of about 50% on this junction, whereas a maximum of 300% was found in some devices.

The result is consistent with the model of the ferroelectric Schottky diode. Indeed, since the work function of Ti is lower than the affinity of the semiconductor (4.33 eV[36] and 4.8 eV[37], respectively), a downward band bending is expected at the interface, resulting in an inward built-in electric field. Thus, the lower (higher) junction resistance corresponds to inward (outward) ferroelectric polarization (inset of Fig. 2a).

Figure 2c shows the cyclability of the junctions checked by applying a sequence of negative and positive saturating pulses. Remarkably, the maximum endurance obtained in those devices was on the order of $10^5$ cycles. Figure 2d reports the modulation of the junction conductance versus writing pulse width, with no major change down to 250 ns, the lowest pulse width accessible with our instrumentation. This suggests that switching times below 100 ns can be foreseen. While in ferroelectric oxides the large capacitance requires miniaturization, the resistive behaviour of metal/GeTe junctions enables fast switching even for large electrodes. These results set a solid ground for reconfigurable GeTe-based devices exploiting ferroelectricity-driven phenomena.

## Ferroelectric control of spin-to-charge conversion

The influence of ferroelectricity on spin-to-charge conversion (SCC) was studied by combining gating and spin pumping in epitaxial Au(3)/Fe(20 nm)/GeTe(15 nm)/Si samples. In a spin pumping (SP) experiment, the magnetization **M** of the FM layer is saturated perpendicularly to the slab (*x* axis of Fig. 3a) by a *dc* magnetic field **H** (Fig. 3a). The precession of **M** is excited by a microwave field **h**$_{rf}$ oscillating along the longitudinal axis *y* at a fixed frequency. The ferromagnetic resonance (FMR) condition is found by sweeping the amplitude of the external saturating field **H**. As a consequence of angular momentum conservation, the additional damping of **M** ascribed to GeTe corresponds to the injection of a *dc* pure spin current **j**$_S$ along *z* into the GeTe layer[38], with a spin polarization **P**$_S$ parallel to the equilibrium magnetization. Spin-to-charge conversion would result in the generation of a charge current **j**$_C$ that is observed in the direction parallel to the longest side of the slab (*y*) as a voltage drop Δ*V* between its edges. The value of the produced charge current density, normalized by the slab width $w$ and $h_{rf}^2$ (proportional to the power), can be expressed as $I_c = (\Delta V/R)/(h_{rf}^2 w)$, where $R$ is the longitudinal resistance of the slab[39].

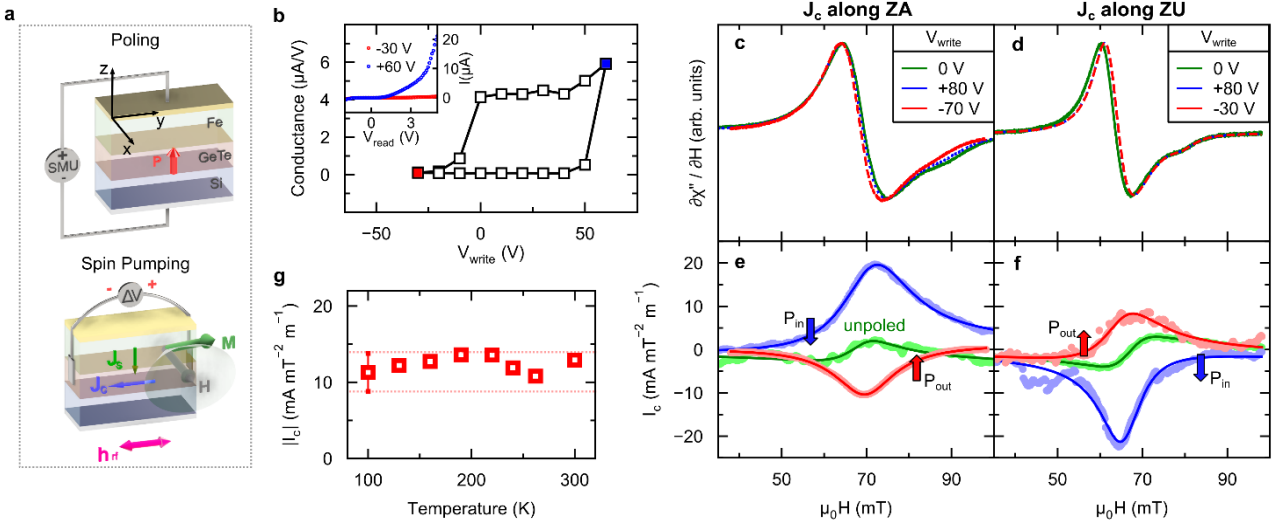

**Figure 3 | Ferroelectric control of the spin-to-charge conversion in GeTe investigated by SP-FMR. a,** Setup for the study of the ferroelectric switching of the spin-charge conversion in GeTe. Above, electrical circuit for ferroelectric switching monitored by resistance changes. Below, the sketch of the contacts used to measure the lateral voltage proportional to the charge current production in the same experiment. Negative (positive) voltage pulses were applied by a source-measure unit (SMU) to stabilize $P_{out}$ ($P_{in}$). **b,** Hysteresis loop of the conductance versus $V_{write}$ of Au(3 nm)/Fe(20 nm)/GeTe(15 nm)/Si. In the inset, $I$-$V$ curves of the heterostructure after the application of two saturating voltage pulses at $V_{write}$= -30 V and +60 V. FMR spectra (**c-d**) and normalized current production (**e-f**) for the slab oriented along $ZA$ and $ZU$ versus ferroelectric polarization. Blue curves correspond to $P_{in}$ ($V_{write}$< 0) and red to $P_{out}$ ($V_{write}$> 0). The spin pumping peak is positive (negative) for $P_{in}$, negative (positive) for $P_{out}$. The green curve in panels e and f refers to the pristine (unpoled) states. The relatively small amplitude of the SP signal in the unpoled state is associated to a multi-domain ferroelectric configuration. **g,** Temperature dependence of the charge current production. The error bar (standard error, SE) reported at 100 K is calculated from a set of measurements.

The ferroelectric polarization was set to be either $P_{in}$ or $P_{out}$ by applying proper voltage pulses, as monitored by measuring the junction resistance. The Au/Fe layer was the gate electrode, while the Si substrate was used as a back contact (cf. Fig. 3a). Fig. 3b shows $I$-$V$ curves at saturation, and the hysteresis loop of the conductance versus $V_{write}$ confirms the ferroelectric switching. Due to the voltage drop across the thick Si substrate, higher voltages (~50 V) are required for such switching. In this configuration, the overall resistance is determined by the silicon substrate, in series with two junctions, Fe/GeTe and GeTe/Si. The latter dominates because of the larger difference in work functions/electron affinities ($\chi_{GeTe}$= 4.8 eV[37], $\chi_{p\text{-}Si(111)}$= 4.0 eV[36], $\phi_{Fe(111)}$≈ 4.81 eV[36]) and the relatively low doping of Si, which determines a wider Schottky barrier. Therefore, the Si/GeTe interface provides an even more effective readout of the memory state, with a modulation of the resistance much larger (~ 4000%) than that of Ti/GeTe.

Fig. 3c-f summarizes the investigation of SCC versus FE polarization. Upon application of voltage pulses, no major differences in shape and position of the FM resonance of the Fe layer were detected (panels c and d), which indicates that the magnetic layer was not affected by the process. The absolute value of the normalized current production $I_C$ is higher in the two saturated states $\mathbf{P}_{out}$ and $\mathbf{P}_{in}$ with respect to the unpoled. Remarkably, the sign of the SCC reverses with the FE polarization (panels e and f), demonstrating the non-volatile ferroelectric control of the spin-to-charge conversion in GeTe. Note that the current production is relatively large, between 10 and 20 mA·mT$^2$·m$^{-1}$. By repeating the poling several times to fully saturate the ferroelectric state, it eventually increases up to 66 mA·mT$^2$·m$^{-1}$, comparable to the 77 mA·mT$^2$·m$^{-1}$ observed with the same setup in platinum (see Ref. [40]), which is one of the reference materials for the generation of spin currents in fundamental studies and applications[41]. Fig. 3g reveals that the amplitude of the SCC in GeTe remains constant from 100 K up to 300 K, demonstrating that GeTe is suitable for tuneable and effective generation of pure spin currents at room temperature in a semiconducting platform compatible with CMOS. Contribution to the signal coming from thermoelectric effects in GeTe could be expected, but thermal gradients are not affected by ferroelectric polarization switching.

Finally, measurements performed along two different crystallographic orientations, corresponding to the high-symmetry directions *ZA* and *ZU* (panels c-e and d-f in Fig. 3, respectively), revealed a charge production of different sign for the same ferroelectric polarization, in accordance with symmetries of the calculated spin-to-charge conversion tensors.

## Discussion

Two mechanisms could cause the observed conversion from spin to charge currents: the inverse Rashba-Edelstein effect (IREE) and the inverse spin Hall effect (ISHE). In IREE, when a spin accumulation is produced by spin pumping in the Rashba bands, the two Fermi contours are displaced producing a net charge current perpendicular to both spin current and spin polarization. The switching of the chirality of spins achieved by reversing the ferroelectric polarization would lead to the reversal of the charge current.

We stress however that IREE is a surface/interface effect, unlikely here due to the possible suppression of surface states of GeTe induced by the deposition of Fe[18], while in our system a spin current can propagate within the semiconducting GeTe film, suggesting rather a bulk-like origin of the observed spin-to-charge conversion. Although we cannot completely rule out IREE, the quantitative arguments discussed below suggest a minor contribution of this phenomenon.

In order to understand the role of ISHE in the charge current production and its ferroelectric switching, we performed analysis based on first principles calculations. The intrinsic spin Hall conductivity (SHC) defined by the standard Kubo's formula is invariant under the operation of inversion which connects two ferroelectric ground states of the *bulk* crystal. Even though a sizable SHE was predicted in GeTe[42,43], it should not be affected by simply reversing the spins in the Rashba bands, without additional modifications in the topology of the electronic states. Here, we considered thin films of crystalline GeTe without standard periodic boundary conditions for bulk GeTe. We simulated a finite slab and revealed that the polarization charges at the surfaces of the film, either positive or negative depending on the ferroelectric polarization direction, affect the electric field inside the crystal. This induces non-trivial bands reconstructions and eventually a difference in the spin Hall conductivities calculated for opposite ferroelectric displacements. Figure 4b reports the calculated element of the SHC tensor ($\sigma_{zx}^{y}$) corresponding to the experimental configuration, with spin polarization along *y* and spin/charge currents along *z*/*x* axes, respectively. It is evident that the electric field introduced by the polarization charge can modulate the SCC, consistently with the experimental change of sign upon the reversal of the ferroelectric polarization (Fig. 3e-f).

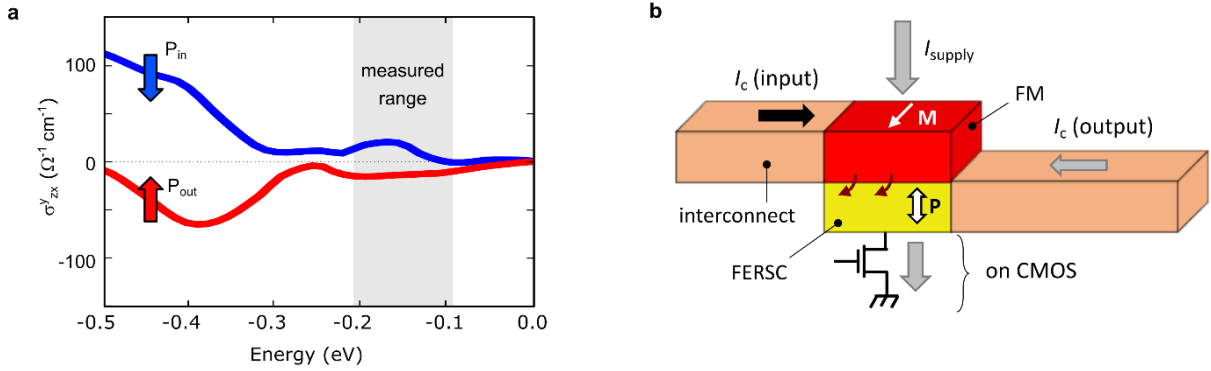

**Figure 4 | Switchable spin to charge conversion in GeTe: theoretical evidence and device concept. a**, Spin Hall conductivity $\sigma^y_{zx}$ calculated for a spin current injected along the (111) direction, with a spin polarization along *ZU* (y) and a charge current along *ZA* (*x*). The switching of the ferroelectric polarization leads to non-trivial modifications of the electronic structure which affect calculated spin Hall conductivities, eventually producing its sign reversal in case of the *p*-type conduction. The grey shaded area identifies the range of energy where conduction occurs. **b**, Concept of the spin-orbit transduction device with FERSC exploiting the ferroelectric control of spin-to-charge conversion. A ferromagnetic layer with a fixed magnetization injects a spin-polarized current in the active FERSC element when the supply current is on. An output charge current is generated by ISHE with a sign related to the polarization state, reversible by an input current (e.g. from another device) providing the required voltage drop across the ferroelectric. The FERSC is both a ferroelectric memory and a spin transducer, eliminating the need for magnetoelectric elements able to drive the magnetization of the spin source. The compatibility with silicon is fundamental towards the integration on CMOS circuitry.

The calculated SHC $\sigma^y_{zx}$ permits to evaluate the efficiency of the spin-to-charge conversion through the spin Hall angle, which is given by $|\theta_{SH}| = |\sigma^y_{zx}|/\sigma^c_{xx}$. The experimental charge conductivity in our films is $\sigma^c_{xx}= 3\times10^3$ $\Omega^{-1}$ cm$^{-1}$ and $\sigma^y_{zx}$ is of the order of 20 $\Omega^{-1}$ cm$^{-1}$ for thin films (gray area in Fig. 4b). While the sign of $\theta_{SH}$ depends on the polarization direction, its absolute value is around 1%, in agreement with similar spin pumping curves obtained for both GeTe and Pt[44,45]. This corroborates a dominant role of the ISHE in the SCC in Fe/GeTe heterostructures.

The ferroelectric control of spin-charge interconversion in FERSC on silicon permits to envisage great advances in beyond-CMOS computing. Exemplarily, the concept device sketched in Fig. 4b simplifies the magnetoelectric spin-orbit element (MESO) proposed by Intel[15]. While the original architecture comprises one magnetoelectric element to drive a ferromagnet and a further spin-orbit interface for the spin transduction, FERSC allows for both memory and spin transduction within the same material, thereby eliminating the need of complex magnetoelectric elements to control the magnetization.

**Conclusion**

In summary, we demonstrated the switchability of the ferroelectric polarization in epitaxial films of GeTe through gating by linking the distribution of ferroelectric domains to the resistance of gate/GeTe junctions. The ferroelectric switching is obtained with low voltages (< 5V) and robust, with switching times below 250 ns and endurance up to $10^5$ cycles. The readout of the state is non-destructive and favoured by the significant resistive window achieved for both metal/GeTe and GeTe/Si interfaces (~300% and ~4000%, respectively).

We exploited the gating of GeTe to reveal the impact of the ferroelectric polarization on the spin-to-charge conversion. By spin pumping, we observed a normalized charge current production in Fe/GeTe at room temperature, whose magnitude is comparable with that of the reference material Pt. Strikingly, the sign of the conversion coefficient can be reversed by acting on the ferroelectric polarization. Whereas theoretical arguments suggest that both Rashba-Edelstein and spin Hall effects could account for the switchable spin-to-charge conversion, the good agreement between the experimental data and the calculated spin Hall angle suggest a dominant role of ISHE in Fe/GeTe.

These achievements lay the foundation for an effective non-volatile electric control of spin currents in the semiconductor GeTe, allowing for monolithic integration on silicon of beyond-CMOS devices and thus paving the way to the realization of reconfigurable spin-based in-memory computing devices.

## Acknowledgements

C.R. acknowledge the financial support by Fondazione Cariplo, grant No. 2017-1622 (ECOS) and by the Italian Ministry of University and Research (MIUR) under the PRIN program, project No. 2017YCTB59 (TWEET. ToWards fErroElectricity in Two-dimensions). M.B. acknowledges support from the European Research Council through the Advanced Grant "FRESCO" #833973. J.S. acknowledges Rosalind Franklin Fellowship from the University of Groningen. We acknowledge the financial support by ANR French National Research Agency Toprise (No. ANR-16-CE24-0017), ANR French National Research Agency OISO (No. ANR-17-CE24-0026), the Laboratoire d'excellence LANEF (No. ANR-10-LABX-51-01), We are grateful to the EPR facilities available at the National TGE RPE facilities (No. IR 3443) and to the High Performance Computing Center at the University of North Texas and the Texas Advanced Computing Center at the University of Texas, Austin. We acknowledge Ariel Brenac, Jean-François Jacquot, Christian Lombard, Serge Gambarelli for their help and advice on the FMR measurement setup. We are grateful to Olivier Klein, Matthieu Jamet, Fu Yu, Thomas Guillet, Marcos Guimarães, Domenico di Sante and Fabio Pezzoli for helpful discussions. This work was partially performed at Polifab, the micro and nanofabrication facility of Politecnico di Milano.

## Author contributions

C.R. conceived the experiment and coordinated the research work with the support of R.B.. S.C. and R.C. grown GeTe/Si samples and did the structural characterization. S.V., L.N, S.P. and A.N. fabricated the devices for electrical characterization and PFM experiments. S.V., C.R. and L.N. did PFM imaging. F.F. performed the measurements of switching speed. S.P., C.R., S.V., D.P., E.A., M. Cantoni and S.P. grew the heterostructures for spin pumping experiments, while S.V., P.N. and L.V. and J-P.A. performed those experiments. J.S. performed the calculations with the support of M.B.N.,

M. Costa and S.Picozzi. C.R., M.B., J.S., and S.V. wrote the manuscript, with fundamental inputs from all the authors.

## Competing interests

The authors declare no competing interests.